# Extending Object-Oriented Languages by Declarative Specifications of Complex Objects using Answer-Set Programming

Johannes Oetsch, Jörg Pührer, and Hans Tompits
*Institut für Informationssysteme, Technische Universität Wien*
{*oetsch,puehrer,tompits*}*@kr.tuwien.ac.at*

*Abstract*—Many applications require complexly structured data objects. Developing new or adapting existing algorithmic solutions for creating such objects can be a non-trivial and costly task if the considered objects are subject to different application-specific constraints. Often, however, it is comparatively easy to declaratively describe the required objects. In this paper, we propose an approach for instantiating objects in standard object-oriented programming languages. In particular, we extend JAVA with declarative specifications in terms of answer-set programming (ASP)—a well-established declarative programming paradigm from the area of logic-based artificial intelligence—from which the required objects can be automatically generated using available ASP solver technology.

## I. INTRODUCTION

Imagine one has to write an algorithm for solving the following problem: Given an array of network components of three different types, where each of the components has potentially multiple cable sockets, create an undirected network graph where each node contains a component. The number of edges incident to a node is limited by the number of sockets of the respective network component. Moreover, the total number of edges must not exceed a given limit, and every component has to be transitively reachable by every other component. Also, there must not be an edge between nodes with components of the same type. The graph is represented by instances of a class Node that store references to their adjacent nodes. Also, we want to identify a node with a maximal number of edges. It will probably require some thought to come up with an algorithm that produces a respective graph structure whenever there exists one. In general, developing an algorithmic solution is sometimes a non-trivial task when data structures need to be generated that are subject to complex constraints and algorithmic off-the-shelf solutions to obtain them are unavailable. Although it might be unclear *how* some desired objects can be created, it is in many cases easy to describe them, i.e., to state *what* is needed, as suggested by the description of the above problem (used as a running example in the remainder). Altough such a situation is unpleasant when one sticks to a traditional imperative programming style, this is the perfect starting point for solving the problem using declarative programming.

Our goal is to integrate declarative object specifications in object-oriented programming languages that allow for obtaining the desired instances automatically. These specifications are especially beneficial in situations where

- **no algorithm is known** for producing the desired objects and the development of a new procedural solution is expensive;
- objects are needed that are **subject to complex constraints**;
- the programmer is confronted with **changing requirements** on the data structures to instantiate;
- **rapid prototyping** is needed, i.e., when an algorithmic solution is the final goal but not feasible before most requirements have been settled.

Our concrete proposal is to combine JAVA with *answer-set programming* (ASP), a well-established paradigm for declarative problem solving from the area of logic-based artificial intelligence (AI). The idea of ASP is to declaratively specify a computational problem in terms of a logical theory (that is, a logic program) such that the solutions of the problem correspond to the models (the "answer sets") of the theory. Following the principle of declarative methods of separating the *representation* from the *processing* of knowledge, ASP solvers are used to compute the models of the specifications. ASP has its roots in nonmonotonic reasoning and allows for expressing *constraints*, *recursive definitions*, and *non-determinism* in a quite natural way. ASP has been used in a wide range of applications, including semantic-web reasoning, systems biology, software testing, linguistics, diagnosis, and, most prominently, for realising a decision support system for the Space Shuttle aiding the pilot in the presence of multiple failures.

Adapted to the setting of this paper, the answer-set program corresponds to a specification of the desired objects, whereas the problem solutions are the objects themselves. ASP solvers allow for exhaustively generating all, a specified number, or a random sample of objects fullfilling the specifications. Also, answer-set programs may contain optimisation expressions that allow to express preferences for certain objects. In the network example, we could search for graphs that

This work has been submitted to the 34th International Conference on Software Engineering, NIER Track, and was partially supported by the Austrian Science Fund (FWF) under project P21698.

involve only a minimal number of edges. Since ASP solvers implement complete search, we can also show that no object meets some specification. In view of the constant improvement of ASP solver technology, the use of ASP for declarative programming becomes increasingly attractive.

In order to avoid programming overhead that would come with calling an ASP solver as an external application within JAVA code, we propose to tightly integrate JAVA elements into the ASP language. This way, on the one hand, JAVA objects, arrays, or primitive data types, that serve as parameters for the specification can be automatically translated to input for a respective solver and, on the other hand, the returned answer sets can be automatically interpreted to build up the desired JAVA data structures. In particular, our proposed formalism allows JAVA constructors, method calls, and objects to take the role of ASP predicates and terms. By means of those, the programmer may specify objects by describing an arrangement of constructor and method calls such that the resulting objects satisfy the constraints of the application.

The interface between the procedural and the declarative parts is realised by a JAVA method call that takes parameters of the specification and the number of desired solutions as arguments. In return, we get a collection of solution objects that meet the specifications. We implemented a proof-of-concept tool for our proposed specification language.

## II. ANSWER-SET PROGRAMMING IN A NUTSHELL

ASP has been proposed as a problem solving approach in the late 1990s, building on the *stable-model semantics* for logic programs [1] that is genuinely declarative—in contrast to, e.g., the semantics of PROLOG. ASP solvers have become increasingly efficient in recent years. In fact, the solver CLASP [2] even outperformed state-of-the-art SAT solvers at the latest SAT competitions in several categories (see http://www.satcompetition.org).

ASP comes with high-level modelling capabilities that allow for specifying problems in an easy-to-read, compact, and elaboration tolerant way, i.e., small variations in a problem description require only small modifications of the representation.

For space reasons, we omit a formal introduction of syntax and semantics of ASP and sketch only its basic ideas; for a comprehensive introduction to ASP, we refer to the well-known textbook by Baral [3].

Roughly speaking, an answer-set program is a collection of rules like `a(X?,Y?):- b(X?,Y?), not c(Y?)`, where `a(X?,Y?)`, `b(X?,Y?)`, and `c(Y?)` are atoms that might be true or false, and `X?` and `Y?` are schematic variables that stand for an object from a domain. Sometimes, the symbol "_" is used to denote a fresh variable not appearing anywhere else. The intuition of the rule is that, for all objects `o1` and `o2`, if `b(o1,o2)` is true and it is not known that `c(o2)` is true, then `a(o1,o2)` must be true. This understanding of the negation operator `not` is called *default*

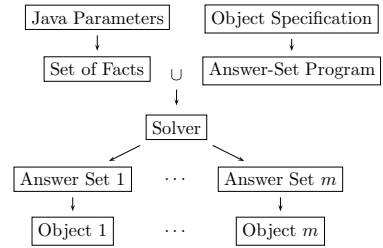

Figure 1. Applying the ASP paradigm for specifying objects.

*negation*, or *negation as failure*, allowing to expressing non-determinism. Consider the following program:

`a(o):- not b(o).  b(o):- not a(o).  c(o):-.`

It has two answer sets, {`a(o),c(o)`} and {`b(o),c(o)`}. For the first, as atom `b(o)` is not known to be true, the first rule is active and derives `a(o)`, whereas the second rule is inactive since `a(o)` is known to be true. Symmetrically, for answer set {`b(o),c(o)`}, the second rule is active but the first one is not. The rule "`c(o):-.`" is a *fact* stating that `c(o)` is unconditionally true. Another type of rules are *constraints* that do not derive anything but are used for eliminating unwanted answer sets. E.g., the constraint "`:-a(o).`" expresses that `a(o)` cannot be true. If added to the program from above it would eliminate the answer set {`a(o),c(o)`}.

We often use special atoms called *cardinality constraints* that allow for reasoning about sets of atoms, e.g., the cardinality constraint `2{edge(X?,Y?):X?<Y?}4` is only true if at least two but at most four edge atoms are true for which the first argument is smaller than the second argument. It is assumed that < is a comparison relation for all objects.

## III. MAIN APPROACH

Figure 1 illustrates the basic idea of our approach for adopting the ASP paradigm for automatic object instantiation. The programmer only provides a declarative specification and parameter values that are automatically translated to an ASP program such that the resulting answer sets are in one-to-one correspondence with all objects satisfying the specification for the given parameters. Depending on the needs of the application, one can compute all or just a predefined number of solutions. Desired objects can then be instantiated automatically from the answer sets. Intuitively, the key for realising such a behaviour is extending the domain over which we reason in ASP to JAVA objects and data values while providing special predicates and function symbols for accessing, creating, and returning JAVA objects and arrays and invoking constructors and object methods in the specification. In the following, these new elements are illustrated by solving the example from the introduction.

Assume network components are instances of class `Component` that has the getter-methods `getNrSock()` and `getType()`, both returning positive integers, where the domain of the latter is from $\{1, 2, 3\}$. Nodes are represented by instances of class `Node`:

```
package example.graph;
public class Node {
 Component c;
 List<Node> nodes = new ArrayList<Node>();
 public Node(Component c){
  this.c=c;
 }
 public void addNode(Node node){
  nodes.add(node);
 }
 ... // getters/setters
}
```

The structure of the specification of the graph is as follows.

```
package example;
import example.graph.*;
NetworkSpec(Component[] comps, int nrCables){
 ... // ASP code
}
```

Similar to regular JAVA class files, a specification belongs to a package and may import classes from other packages as needed. In the following, the missing ASP code is introduced and explained step-by-step.

The first rule is a fact that consists of a cardinality constraint. It defines the search space by guessing whether there is an edge between any two different components.

```
0 {edge(C1?,C2?) : C1? != C2? :
               C1?comps(_) : C2?comps(_)} 1.
```

The first type of expression that extends standard ASP for use with JAVA is of form `C?comps(I?)`, a special atom that is true if the variable `C?` is assigned the object with index `I?` in the array contained in specification parameter `comps`. Consequently, `edge` is a relation between the `Component` objects from the input array.

Next, since we deal with an undirected graph, the following rules (in pure ASP) ensure the symmetry of edges and transitively compute reachability between components.

```
edge(C1?,C2?):- edge(C2?,C1?).
reach(C1?,C2?):- edge(C1?,C2?).
reach(C1?,C2?):- reach(C1?,H?),reach(H?,C2?).
```

For ensuring that the number of edges from a component exceeds its number of sockets, the next constraint is added.

```
:- C1?.getNrSock()+1 {edge(C1?,C2?):
                  C2?comps(_)}, C1?comps(_).
```

The term `C1?.getNrSock()` stands for the value returned by the method `getNrSock()` of the object contained in variable `C1?`. The intuitive reading of the entire rule is that for any object `C1?` with an arbitrary index in parameter array `comps` it cannot hold that the number of edges to other objects `C2?` in `comps` is greater or equal to the number of sockets of `C1?` plus 1.

Next, the number of edges is restricted to the value of the integer parameter `nrCables`.

```
:- nrCables+1 {edge(C1?,C2?) : C1? < C2? :
              C1?comps(_) : C2?comps(_)}.
```

The final constraints ensure that every component is reached from every other and that there is no edge between components of same type.

```
:- C1?comps(_), C2?comps(_), C1? != C2?,
                   not reach(C1?,C2?).
:- edge(C1?,C2?), C1?comps(_), C2?comps(_),
        C1?.getNrSock() == C2?.getNrSock().
```

From a logical point of view the problem is solved here. What remains is the declarative specification of how `Node` objects should be instantiated and configured and to determine a return value.

```
new Node(C1?):- C1?comps(_).
exe N1?.addNode(N2?):- N1?Node(C1?),
                  N2?Node(C2?),edge(C1?,C2?).
```

For every `Component` object `C1?`, we derive an atom `new Node(C1?)` representing a respective constructor call with the component as argument, and hence the instantiation of a `Node` object containing the component. Objects created this way can be referenced in other rules in a similar fashion as the elements of the `comps` array. In particular, an atom `N1?Node(C1?)` is true if `N1?` is the object created by constructor call `new Node(C1?)`. The last rule states that for two nodes `N1?` and `N2?` whose components are connected by an edge, the `addNode(Node node)` method of `N1?` should be invoked with `N2?` as argument. This is expressed by the use of special atom `exe N1?.addNode(N2?)` representing a method invocation that is automatically executed after the specified objects have been created.

It is required that one of the created nodes is returned that has the highest number of edges. As there might be several, we choose the minimal according to the order <.

```
nrEdges(C1?,Nr?):- Nr? = {edge(C1?, C2?) :
             C2?comps( _)}, C1?comps(_).
notReturn(C1?):- nrEdges(C1?,Nr1?),
           Nr1? < Nr2?, nrEdges(C2?,Nr2?).
notReturn(C1?):- nrEdges(C1?,Nr?),
           C1? > C2?, nrEdges(C2?,Nr?).
return N?:- N?Node(C?), not notReturn(C?).
```

This completes the rules that suffice to describe the desired objects from the example problem. Next, we show how to access the specification from JAVA code to obtain the specified graph.

```
Component[] comps = {c1,c2,c3,c4,c5,c6};
NetworkSpec spec = new NetworkSpec();
spec.evaluate(comps,9, 1);
if(spec.hasSolution()){
 Node res = (Node)spec.getSolutions().get(0);
```

First, the array `comps` is created that contains six components assumed to be initialised earlier. Then, an object respresenting the `NetworkSpec` specification is instantiated. We call its method `evaluate` that takes as arguments the parameters of the specification, `comps` and `nrCables`, and an additional `int` parameter determining the number of desired solutions, here 1 (0 stands for all). The `hasSolution` method checks whether a desired graph exists, which might not be the case for some amounts of sockets and types of components. If one exists, it is assigned to the variable `res`. Note that the problem could be solved in only 23 lines of combined ASP and JAVA code.

There is also the possibility to employ optimisation statements for ASP solvers, e.g., adding the statement "`#minimize{edge(C1?,C2?)}.`" to the specification allows for searching solutions with a minimal number of edges. Moreover, if random objects are needed, techniques from SAT can be used to get a near uniformly distributed selection from the set of all specified objects [4].

We developed a prototype implementation of our approach which translates a specification into a JAVA class (the `NetworkSpec` class in the example) that at runtime uses CLASP as an external solver, converts given parameters into facts, and instantiates and configures the desired objects according to the obtained answer sets. Here, JAVA objects are internally represented by automatically generated identifiers. Features that are already supported but not part of the example is the creation of arrays from specified or parameters objects, nested constructor and method calls, and the possibility to specify an order of execution for calls using an `exe` atom.

## IV. Related Work

Constraint programming (CP) is a declarative programming paradigm often used by imperative languages through respective CP libraries. Typically, constraints are formulated for variables over primitive data type domains. The embedding in object-oriented languages is mostly realised by wrapper classes for variables, constraints, and solvers. Output in CP is given as vectors of variable assignments satisfying the constraints, which is opposed to structured information in answer sets that we exploit for building up complex objects.

Declarative specification of complex structures is often used for testing. E.g., ALLOY is a declarative first-order language for specifying objects for bounded-exhaustive testing, i.e., all objects that do not exceed a given size are used as test input for a piece of code [5]. The TestEra framework provides means to translate ALLOY instances to JAVA objects [6]. As ALLOY structures are generated offline, specifications do not allow for runtime parameters as in our approach, other than size limitations. Moreover, as the target is to consider all solutions up to some size, there is no support for getting optimal ones. Another JAVA test input generator is KORAT, which is, however, based on procedural specifications [7]. Object trees are generated as solution candidates which are then checked against a checking method that accepts when the structure is a solution. Hence, individual constraints cannot help to prune the search space. Indeed, we see testing as one application for our combined language, e.g., red-black trees can be concisely specified in our approach that are a popular example for test input generation in the testing literature.

## V. Future Prospects

There are many ways to continue work from here. For one, the current specification language still misses important JAVA features, like direct field assignments and static method calls that could be easily integrated.

Currently, when the result of a call to an object method needs to be considered for solving, like when using `C1?.getNrSock()`, the objects considered for `C1?` has to exist before solving. Moreover, in our current implementation, we need to state that `getNrSock()` has to be precomputed for all objects in the `comps` array. The reason is that potential values for `C1?` are determined only during solving. Here, allowing the solver to interact with the JAVA-runtime can help. In this respect, a tight integration of a solver and a virtual machine would be desirable as this not only allows for executing JAVA code during solving, but also for exploiting the same data structure, e.g., pointers of objects as their identifiers, and reducing overhead for external calls.

Another possibility is to develop a tool that translates specifications without parameters into JAVA code generating the specified objects without calling a solver at runtime.

In conclusion, we see a high potential for significantly reducing the effort that has to be spend in a software project for writing and testing involved imperative code by integrating declarative specifications.

## References

[1] M. Gelfond and V. Lifschitz, "The stable model semantics for logic programming," in *Proc. ICLP'88*, 1988, pp. 1070–1080.

[2] "Potassco—the Potsdam answer set solving collection," http://potassco.sourceforge.net, last visited: Nov. 30, 2011.

[3] C. Baral, *Knowledge Representation, Reasoning, and Declarative Problem Solving*. Cambridge, England, UK: Cambridge University Press, 2003.

[4] C. P. Gomes, A. Sabharwal, and B. Selman, "Near-uniform sampling of combinatorial spaces using XOR constraints," in *Proc. NIPS'06*. MIT Press, 2006, pp. 481–488.

[5] D. Jackson, "Alloy: A lightweight object modelling notation," *ACM Trans. Softw. Eng. Methodol.*, vol. 11, pp. 256–290, 2002.

[6] S. Khurshid and D. Marinov, "TestEra: Specification-based testing of Java programs using SAT," *Autom. Softw. Eng.*, vol. 11, no. 4, pp. 403–434, 2004.

[7] A. Milicevic, S. Misailovic, D. Marinov, and S. Khurshid, "Korat: A tool for generating structurally complex test inputs," in *Proc. ICSE'07*. IEEE Computer Society, 2007, pp. 771–774.